%% file: main.tex
\documentclass[conference,a4paper]{IEEEtran}

\renewcommand\IEEEkeywordsname{Index Terms}

\usepackage{amsmath,amssymb,bm}
\usepackage{graphicx}
\graphicspath{{fig/}{result_checklist_figures/}}
\usepackage{booktabs}
\usepackage{multirow}
\usepackage{cite}
\usepackage[hyphens]{url}
\usepackage[bookmarks=false]{hyperref}
\usepackage[nolist]{acronym}
\usepackage{xcolor}
\input{corporateColours}
\usepackage{float}
\floatstyle{ruled}
\newfloat{algorithm}{t}{loa}
\floatname{algorithm}{Algorithm}

\usepackage{tabularx}

\usepackage{tikz}
\usetikzlibrary{arrows.meta,positioning,calc,fit,patterns}

\usepackage{pgfplots}
\pgfplotsset{compat=1.18}
\usepgfplotslibrary{groupplots}
\usepgfplotslibrary{fillbetween}

\newcommand{\Hm}{\mathbf{H}}

\newcommand{\cv}{\mathbf{c}}

\newcommand{\sv}{\mathbf{s}}
\newcommand{\Lv}{\mathbf{L}}
\newcommand{\Ftwo}{\mathbb{F}_2}
\newcommand{\Ccal}{\mathcal{C}}
\newcommand{\Lcal}{\mathcal{L}}
\newcommand{\Bcal}{\mathcal{B}}
\newcommand{\Ncal}{\mathcal{N}}
\newcommand{\Rcal}{\mathcal{R}}
\newcommand{\Fcal}{\mathcal{F}}
\newcommand{\Mcal}{\mathcal{M}}

\newcommand{\Ecal}{\mathcal{E}}
\newcommand{\ind}[1]{\mathbf{1}\!\left[#1\right]}

\IEEEoverridecommandlockouts
\begin{document}
\setlength{\columnsep}{0.21in}
\bstctlcite{IEEEexample:BSTcontrol}

\title{Row-Boosted Ensemble Belief Propagation for Short LDPC Codes}

\author{\IEEEauthorblockN{Paul Bezner\textsuperscript{\dag}, Felix Krieg\textsuperscript{\dag}, and Stephan ten Brink}
\IEEEauthorblockA{Institute of Telecommunications, University of Stuttgart, Germany\\
\{bezner, krieg, tenbrink\}@inue.uni-stuttgart.de}
\thanks{\textsuperscript{\dag}\,P. Bezner and F. Krieg contributed equally to this work.}
\thanks{This work is supported by the German Federal Ministry of Research, Technology and Space (BMFTR) within the project Open6GHub+ (grant no. 16KIS2406).}
\thanks{LLM agents from Anthropic and OpenAI were used to write source code for this work and to assist with language editing.}}

\maketitle

\begin{acronym}
\acro{AED}{automorphism ensemble decoding}
\acro{AWGN}{additive white Gaussian noise}
\acro{BER}{bit error rate}
\acro{BLER}{block error rate}
\acro{BP}{belief propagation}
\acro{BPSK}{binary phase-shift keying}
\acro{CN}{check node}
\acro{FER}{frame error rate}
\acro{LDPC}{low-density parity-check}
\acro{LLR}{log-likelihood ratio}
\acro{MBBP}{multiple-bases belief propagation}
\acro{ML}{maximum likelihood}
\acro{MRB}{most-reliable basis}
\acro{MSA}{min-sum algorithm}
\acro{NMSA}{normalized min-sum algorithm}
\acro{OSD}{ordered-statistics decoding}
\acro{PCM}{parity-check matrix}
\acro{RBE}{row-boosted ensemble}
\acro{SMS}{saturated min-sum}
\acro{SPA}{sum-product algorithm}
\acro{SNR}{signal-to-noise ratio}
\acro{VN}{variable node}
\acro{QC}{quasi-cyclic}
\end{acronym}

\begin{abstract}
Belief-propagation (BP) decoding of short and moderate-length low-density parity-check (LDPC) codes is limited by finite-length graph effects: a single decoder trajectory can become trapped or oscillatory even when an alternative trajectory would decode the received word. Existing ensemble-BP decoders create the required diversity through multiple parity-check matrices, automorphisms, modified schedules, subcodes, or altered update rules. We introduce row-boosted ensemble (RBE) decoding as a minimal decoder-side diversity mechanism: all ensemble members share the same parity-check matrix and the same BP kernel, and differ only in a small set of parity-check rows whose outgoing messages are boosted. On the 5G~NR BG1 \((144,96)\) code, RBE with \(32\) members lowers the frame-error rate of BP with \(20\) iterations (BP-20) from \(1.6\times10^{-2}\) to \(1.6\times10^{-3}\) at \(E_\mathrm{b}/N_0=4.0\,\mathrm{dB}\), outperforming saturated-min-sum and affine subcode ensembles of equal size. Increasing the ensemble size yields additional gains, indicating that RBE provides a scalable performance–complexity tradeoff. The gains transfer across 5G~NR block lengths and rates, to non-5G short LDPC codes, and across flooding and layered schedules.
\end{abstract}

\acresetall

\section{Introduction}
\label{sec:introduction}

\Ac{LDPC} codes are decoded efficiently by sparse-graph message passing and are part of modern communication standards. At short and moderate block lengths, however, a single \ac{BP} decoder remains far from \ac{ML} performance. Failures are dominated by the interaction between the received word and finite-length graph structures: short cycles, trapping and absorbing sets, and oscillatory message trajectories~\cite{Richardson03error-floorsof,shin2007trajectory}. Raising the iteration limit helps only when the underlying trajectory eventually escapes on its own; a single decoder has no way to handle error patterns to which it is inherently unsuited -- it lacks \emph{decoder diversity}.

Ensemble-\ac{BP} decoding removes this limitation by running several related decoders in parallel and selecting a valid candidate according to the channel metric. Diversity can be obtained from multiple parity-check bases (\ac{MBBP})~\cite{Huber,hehn2010multiple}, code automorphisms or broken-graph transforms~\cite{Chen_Cyclic_LDPC_AED,geiselhart2022brokengraph}, affine subcodes~\cite{mandelbaum2026affinesubcode}, message saturation~\cite{LDPC_errorFloor_LLRClipping_3}, or schedule variation~\cite{CRC_BPL_ISIT20,krieg2025ensemble}, to name a few. These mechanisms are effective, but each requires an additional ingredient -- such as a second parity-check representation or a non-absorbed code symmetry -- that is not always available~\cite{krieg2025ensemble}.

\begin{figure}[t]
    \centering
    \makebox[\linewidth][c]{%
        \resizebox{0.96\linewidth}{!}{%
            \input{fig/_rbe_dec_block}%
        }%
    }
    \vspace{-0.4cm}
    \caption{Row-boosted ensemble decoder. Every member uses the original matrix \(\Hm\) and the same BP kernel; only the row-boost configuration \(\theta_e=(\Bcal_e,\beta_e)\) differs. The \ac{ML} selection step selects the codeword estimate with the highest likelihood from the list $\Lcal$.}
    \label{fig:rbe_architecture}
    \vspace{-1.5em}
\end{figure}
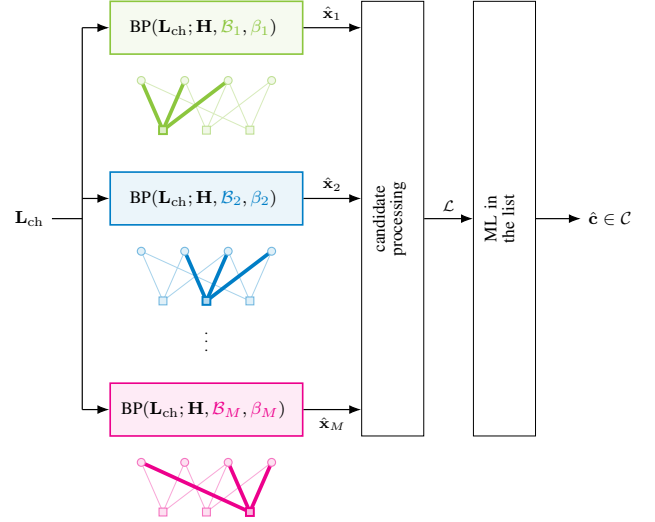

This paper introduces \ac{RBE}-\ac{BP} decoding as a code-agnostic ensemble method (Fig.~\ref{fig:rbe_architecture}). Each ensemble member uses the original \ac{PCM} and the same \ac{BP} kernel, but boosts the outgoing messages of a different small subset of parity-check rows. \ac{RBE} decoding thereby alters the iterative trajectory while preserving the code and the Tanner-graph storage, and can be added to any \ac{SPA}, \ac{MSA}, \ac{NMSA}, flooding, or layered implementation. \Ac{RBE} is related to weighted and neural \ac{BP}, which generalize the message-passing update with trainable edge or iteration weights~\cite{nachmani2018deep}. However, such approaches typically optimize a single weighted decoder and store fine-grained weights across many edges and iterations, whereas \ac{RBE} forms an ensemble from simple row-wise boosts of the original \ac{PCM}.

The contributions are:
\begin{itemize}
    \item We define \ac{RBE} as per-row check-message boosting applied independently to each ensemble member.%
    \item We give a low-effort offline design procedure based on a boost-factor sweep and rescue-rate-based boosted-row selection over a corpus of plain-BP failures, and show that the useful boost factor spans a broad range. %
    \item We analyze which boosted rows lead to favorable amplification and find that a boost is most effective when the support of the row contains no channel errors, i.e., the boosted check acts as a reliable anchor. %
    \item We report \ac{FER}, ensemble-size scaling, decoder-architecture generality (flooding and layered \ac{NMSA}), and cross-code transfer across 5G~NR and other \ac{LDPC} benchmarks, and compare against \ac{MBBP}, broken-graph \ac{AED}, saturated min-sum, subcode, and layered schedule-diversity ensembles of equal size. %
\end{itemize}

\section{Preliminaries}
\label{sec:background}

\subsection{Code and Channel Model}
\label{subsec:code_channel}

An \ac{LDPC} code is defined by a sparse parity-check matrix \(\Hm\in\Ftwo^{m\times n}\),
\begin{equation}
    \Ccal=\{\cv\in\Ftwo^n:\Hm\cv^{\mathsf T}=\mathbf{0}\}.
\end{equation}
The code \(\Ccal\) is thus the set of all length-\(n\) binary vectors that satisfy the \(m\) parity checks given by the rows of \(\Hm\), and has dimension \(k=n-\operatorname{rank}(\Hm)\). Each row of \(\Hm\) defines one parity check, represented by a \ac{CN} in the Tanner graph; we therefore use the terms \emph{row} and \emph{check} interchangeably. A code of blocklength \(n\), dimension \(k\), and rate \(R=k/n\) is denoted by \((n,k)\).
We use \ac{BPSK} over the \ac{AWGN} channel. A bit \(c_j\in\{0,1\}\) is mapped to \(x_j=1-2c_j\), and the received value is \(y_j=x_j+z_j\) with \(z_j\sim\mathcal{N}(0,\sigma^2)\). The channel \ac{LLR} is \(L_{\mathrm{ch},j}=\log\frac{P(y_j|c_j=0)}{P(y_j|c_j=1)}=2y_j/\sigma^2\), collected in the vector \(\Lv_{\mathrm{ch}}\). All simulations use the all-zero codeword, which is valid for the symmetric channel and the symmetric message-passing decoders considered here. For the rate-matched 5G~NR codes, punctured variables are initialized with zero \ac{LLR}, shortened variables with a known saturated value.

\subsection{BP Kernels and Schedules}
\label{subsec:bp_kernels}

For flooding \ac{SPA}, the \ac{VN}-to-\ac{CN} and \ac{CN}-to-\ac{VN} messages are
\begin{align}
    q_{j\to i}^{(t)} &= L_{\mathrm{ch},j}+\sum_{i'\in\Ncal(j)\setminus\{i\}} r_{i'\to j}^{(t-1)},\text{ and}\label{eq:vn_update}\\
    r_{i\to j}^{(t)} &= 2\tanh^{-1}\!\left(\prod_{j'\in\Ncal(i)\setminus\{j\}}\tanh\left(\tfrac{q_{j'\to i}^{(t)}}{2}\right)\right),\label{eq:spa_update}
\end{align}
respectively, where \(\Ncal(j)\) and \(\Ncal(i)\) denote the neighborhoods of \ac{VN} \(j\) and \ac{CN} \(i\) in the Tanner graph.
The \ac{MSA} approximates \eqref{eq:spa_update} by
\begin{equation}
    r_{i\to j}^{\mathrm{MS},(t)}=
    \min_{j'\in\Ncal(i)\setminus\{j\}} \big|q_{j'\to i}^{(t)}\big| \!\prod_{j'\in\Ncal(i)\setminus\{j\}}\!\operatorname{sgn} q_{j'\to i}^{(t)},
\end{equation}
and \ac{NMSA} applies the normalization \(r_{i\to j}^{\mathrm{NMS},(t)}=\alpha\,r_{i\to j}^{\mathrm{MS},(t)}\), with $\alpha\in\mathbb{R}^{+}$. We consider both flooding and layered~\cite{hocevarlayered} schedules with syndrome-based early termination and default to $\alpha=0.75$. An \ac{RBE} scales the \ac{CN} output and is thus independent of these choices.

\subsection{Ensemble Decoding}
\label{subsec:ensemble_dec}

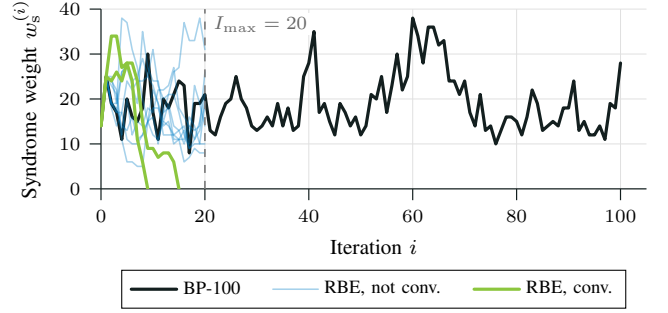
\begin{figure}[t]
    \centering
    \resizebox{0.98\linewidth}{!}{%
        \input{fig/_bg1_n144_k96_rowboost_trajectory_example.tikz}%
    }
    \caption{Syndrome-weight trajectories on a BP-20-failure frame for
    \(\Ccal_\mathrm{5G,BG1}(144,96)\) at \(E_\mathrm{b}/N_0=4.5\,\mathrm{dB}\).
    Plain BP-100 does not converge, while two of ten boosted BP-20 members reach zero syndrome.}
    \label{fig:rowboost_trajectory_example}
    \vspace{-1.5em}
\end{figure}

An ensemble decoder runs \(E\) related \ac{BP} decoders \(D_1,\ldots,D_E\) on the same channel observation \(\Lv_{\mathrm{ch}}\). If the trajectory of one decoder fails, a related decoder may still converge on the same channel observation; Fig.~\ref{fig:rowboost_trajectory_example} shows such an example.

Member \(e\in\{1,\ldots,E\}\) is specified by a configuration \(\theta_e\) -- an alternative parity-check representation, a graph automorphism, a perturbed update rule, or a different schedule -- and returns an estimate \(\hat\cv_e=D_e(\Lv_{\mathrm{ch}};\theta_e)\) after at most \(I_{\max}\) iterations. A member estimate is \emph{valid} if it satisfies all parity checks, i.e., if the syndrome \(\sv:=\Hm\hat\cv_e^{\mathsf T}\) equals \(\mathbf{0}\), making \(\hat\cv_e\) a codeword estimate.  The ensemble collects the valid decoder outputs in a candidate list
\begin{equation}
    \Lcal=\big\{\hat\cv_e : e\in\{1,\ldots,E\},\ \Hm\hat\cv_e^{\mathsf T}=\mathbf{0}\big\}.
    \label{eq:ensemble_list}
\end{equation}
If \(\Lcal=\emptyset\), the ensemble declares a frame error (or outputs a random codeword); otherwise it applies \ac{ML} selection within the list, choosing the candidate that minimizes the frame error probability. As \ac{BP} decoders do not always converge to a valid codeword, post-processing (e.g., \ac{OSD}~\cite{fossorier1995osd} or systematic re-encoding) can ensure that all ensemble members produce valid codewords. This optional post-processing leads to non-empty lists $\Lcal$ and typically increases the decoding performance of the ensemble.

Throughout, BP-\(I\) denotes a plain \ac{BP} decoder with \(I_{\max}=I\); ensembles are labeled by their type and size, e.g., RBE-E32 for an \ac{RBE} with \(E=32\) members.

Concretely, \ac{MBBP} builds an ensemble of algebraically equivalent parity-check matrices, so that cycles and trapping sets tied to one representation are not present in another~\cite{Huber,hehn2010multiple}. 

Automorphism ensemble decoding instead uses code automorphisms to build an ensemble. For codes with a known symmetry (e.g., the cyclic shifts of a cyclic code), each member decodes a permuted copy of \(\Lv_{\mathrm{ch}}\) and maps the result back through the inverse permutation, also leading to a potentially favorable noise representation in some ensemble members~\cite{Chen_Cyclic_LDPC_AED,polar_aed}. In case \ac{BP} is invariant to known automorphisms, the Tanner-graph symmetry can be broken in order to benefit from those automorphisms~\cite{geiselhart2022brokengraph}.

Affine-subcode ensembles add a small set of additional linear constraints to each member's parity-check matrix. Each member decodes a subcode \(\Ccal_e\) whose graph may be better conditioned than that of \(\Ccal\) itself~\cite{mandelbaum2026affinesubcode}. 

Saturated min-sum clips the magnitude of the least reliable message bits and enumerates all possible bit combinations in those positions. Each ensemble member corresponds to one possible bit sequence~\cite{wehn2016afterburner}. 

Schedule-diversity ensembles vary only the check-to-variable update order -- i.e., the order in which the decoder schedules the layered updates~\cite{krieg2025ensemble}.

\section{Row-Boosted Ensemble Decoding}
\label{sec:rbe}

\subsection{Row-Boosting Rule}
\label{subsec:row_boosting_rule}

In an \ac{RBE}, decoder \(D_e\) is specified by a boosted-row set \(\Bcal_e\subseteq\{1,\ldots,m\}\) and a boost factor \(\beta_e\geq0\). After the standard \ac{CN} update, \ac{RBE}-BP scales the contribution of each boosted row to the \ac{VN} posterior,
\begin{equation}
    \tilde r_{i\to j}^{(t,e)}=
    \begin{cases}
        \beta_e\, r_{i\to j}^{(t,e)}, & i\in\Bcal_e,\\
        r_{i\to j}^{(t,e)}, & i\notin\Bcal_e.
    \end{cases}
    \label{eq:rbe_rule}
\end{equation}
All members decode the same \ac{LLR} vector using the same matrix, \(\Hm\); the decoding graph remains unchanged, with only the decoding trajectory differing. With an integer boost factor, \eqref{eq:rbe_rule} is exactly equivalent to replicating each boosted row \(\beta_e\) times in the Tanner graph; the posterior counts the row \(\beta_e\) times. Therefore, boosting requires neither additional matrix storage nor floating-point multipliers for this special case, leading to a hardware-friendly implementation.

\subsection{Decoding Complexity}
\label{subsec:implementation_overhead}

To reduce the decoding complexity of the ensemble, a \emph{gate decoder} can check whether the full ensemble is required~\cite{wehn2016afterburner}. A plain \ac{BP} decoder runs on the unmodified Tanner graph, and the boosted members are only evaluated when the gate produces no valid codeword. The final output is the \ac{ML}-in-list candidate among the ensemble members. This keeps the average decoding complexity close to that of one \ac{BP} decode, since most frames terminate in the gate. In rare cases where the gate converges on a wrong codeword, the ensemble loses the opportunity to ``rescue'' the frame. Therefore, the average complexity of a gated ensemble version is lower, but the \ac{FER} is slightly increased.
\Ac{RBE} adds only a row-boost table to an existing \ac{BP} decoder, thus the average decoding complexity per member is approximately equal to that of a plain \ac{BP} decoder, especially if \ac{NMSA} is used.

\subsection{Offline Boost-Factor Selection}
\label{subsec:beta_sweep}

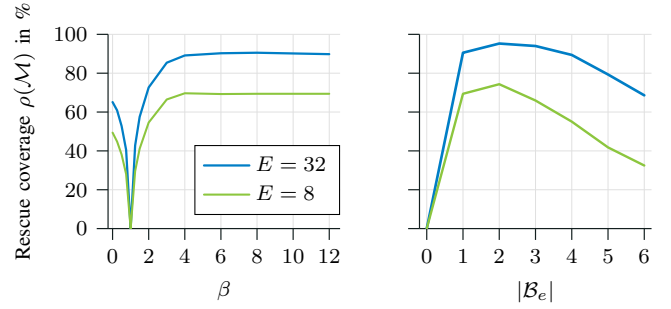
\begin{figure}[t]
    \centering
    \input{fig/fig2_tuning_selection.tikz}
    \caption{Sensitivity of the offline design to the ensemble design parameters for \(\Ccal_\mathrm{5G,BG1}(144,96)\). Left: rescue coverage as a function of the boost factor \(\beta\). Right: coverage versus number of boosted rows \(|\Bcal_e|\).}
    \label{fig:rbe_tuning}
    \vspace{-1.5em}
\end{figure}

For a fixed set of boosted rows \(\{\Bcal_e:e\in\{1,\ldots,E\}\}\), the boost factor \(\beta_e\) controls how strongly the corresponding parity checks perturb the \ac{BP} trajectory and thus directly influences the ensemble performance.
We therefore optimize \(\beta_e\) for a fixed ensemble construction.
To reduce the search space and the implementation overhead, we further constrain the boost factor to be ensemble-wide, i.e., \(\beta_e=\beta\) for all \(e\); empirically, this constraint has a negligible impact on performance.
Specifically, we first construct ensembles of single-row members, i.e., \(|\Bcal_e|=1\), and then sweep \(\beta\) starting from \(\beta=0\). The case \(\beta=0\) removes the contribution of the selected row entirely and can be interpreted as a row-level graph perturbation related to the broken-graph decoder of~\cite{geiselhart2022brokengraph}, whereas \(\beta=1\) corresponds to the unmodified \ac{BP} update and therefore provides no decoder diversity.
The resulting rescue coverage as a function of \(\beta\) is shown in Fig.~\ref{fig:rbe_tuning}~(left).
For both ensemble sizes \(E=8\) and \(E=32\), the coverage saturates for \(\beta\gtrsim 4\) and collapses only at the singular point \(\beta=1\), where all members coincide. Attenuation (\(\beta<1\)) recovers part of the coverage but remains below the maximum attained by strong boosting.
Based on this observation, we set \(\beta=8\) as the default for the row selection procedure.

\subsection{Offline Row Selection}
\label{subsec:validation_corpus}

Each ensemble member $e$ is defined by its set of boosted rows $\Bcal_e$. While the performance of an individual decoder depends only on $\Bcal_e$, the ensemble performance depends on the entire collection of boosted-row sets $\left\{\Bcal_i : i\in\left\{1,\ldots,E\right\}\right\}$. For a given matrix $\Hm$, the best individual member is therefore not necessarily part of the best possible ensemble. Since a joint optimization of the boosted rows over the complete ensemble is computationally demanding, we resort to a sub-optimal selection and build ensembles from the individual candidates with the highest rescue rate relative to the non-boosted \ac{BP} decoder.
For gated \ac{RBE} ensembles, the rescue rate is evaluated on frame samples failed by the plain \ac{BP} gate,
\begin{equation}
    \Fcal=\{\Lv_{\mathrm{ch}}:\hat\cv_{\mathrm{BP}}(\Lv_{\mathrm{ch}})\neq\cv\}.
\end{equation}
Ensemble member $e$ with parameters \((\Bcal_e,\beta_e)\) rescues the set
\begin{equation}
    \Rcal_e=\{\Lv_{\mathrm{ch}}\in\Fcal:D_e(\Lv_{\mathrm{ch}})=\cv\},
\end{equation}
with rescue rate \(\vert\Rcal_e\vert/\vert\Fcal\vert\), and the empirical coverage of an ensemble \(\Mcal\subseteq\{1,\ldots,E\}\) is
\begin{equation}
    \rho(\Mcal)=\tfrac{1}{\vert\Fcal\vert}\Big\vert\textstyle\bigcup_{e\in\Mcal}\Rcal_e\Big\vert.
    \label{eq:coverage}
\end{equation}

In order to reduce the search space, we identify the relevant cardinalities of the set of boosted rows, $\Bcal_e$.
For this purpose, we fix \(\beta_e=8\) and vary the cardinality, constructing ensembles whose members all use the same number of boosted rows \(|\Bcal_e| = \mathrm{const}\).
The resulting rescue coverage as a function of \(|\Bcal_e|\) is shown in Fig.~\ref{fig:rbe_tuning}~(right).
The case \(|\Bcal_e|=0\) corresponds to the unmodified \ac{BP} decoder and therefore provides no additional ensemble gain.
The coverage reaches its maximum at \(|\Bcal_e|=2\) and decreases for larger sets of boosted rows, indicating that overly broad perturbations are less effective for this (short) code.
Based on this observation, we restrict the default search to \(|\Bcal_e|\leq 3\).

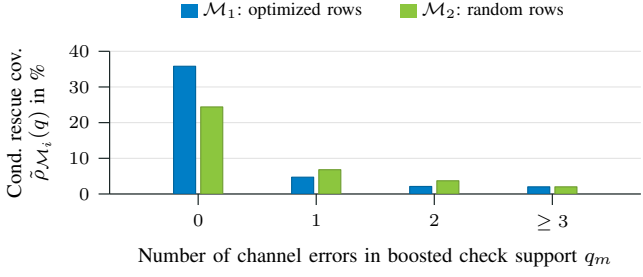
\begin{figure}[t]
    \centering
    \resizebox{0.98\linewidth}{!}{%
        \input{fig/rbe_anchor_rescuE_by_error_count.tikz}%
    }
    \caption{Rescue coverage versus channel errors in the boosted row on BP-20-failure frames for \(\Ccal_\mathrm{5G,BG1}(144,96)\). A single boosted row is most
    effective when its support is error-free.}
    \label{fig:mechanism_rescue}
\end{figure}

\subsection{Boosted Rows as Reliable Anchors and Diversity Enablers}
\label{subsec:mechanism}

Next, we investigate if and when row boosts rescue a given \ac{BP} failure.
Fig.~\ref{fig:rowboost_trajectory_example} shows the decoding trajectories of a single \ac{BP} decoder and of all component decoders of an \ac{RBE}. As a low-complexity indicator of the decoder state and convergence behavior, we use the syndrome weight \(w_s^{(t)}=\sum_i s_i\) at decoder iteration \(t\)~\cite{shin2007trajectory}.
Iteration zero corresponds to the syndrome of the hard decision on the channel observation before decoding starts.
Plain BP-20 and even BP-100 do not reach zero syndrome, whereas two out of ten \ac{RBE} members converge within the iteration limit for this noise sample.
Notably, the set of converging members depends on the channel output, which motivates a closer look at the circumstances under which a row boost rescues a frame.

To determine which perturbations are helpful, we conducted the following investigation: 
For a single-row boost \(\Bcal_e=\{i\}\), let \(q_i=|\Ncal(i)\cap\Ecal_{\mathrm{ch}}|\) count the channel hard-decision errors \(\Ecal_{\mathrm{ch}}=\{j:\ind{L_{\mathrm{ch},j}<0}\neq c_j\}\) that lie in the support of row \(i\).
Grouping all single-row boosts by \(q_i\), we compute the conditional rescue coverage \(\tilde{\rho}(q)\) as the fraction of boosts \(\Bcal_e=\{i\}\) with \(q_i=q\) that recover \(\cv\).
Fig.~\ref{fig:mechanism_rescue} plots \(\tilde{\rho}(q)\) over \(q\), i.e., how often a boosted row rescues a frame as a function of the number of errors in its support, for ensembles built from randomly chosen and optimized rows.

Fig.~\ref{fig:mechanism_rescue} shows that rescue is by far most likely for \(q_i=0\):
an error-free boosted row acts as a reliable anchor that increases the influence of a locally correct neighborhood.
Notably, boosting a row that is directly connected to erroneous \acp{VN} yields a lower but nonetheless non-zero coverage.
The optimized rows selected by the offline procedure of Sec.~\ref{subsec:validation_corpus} exhibit the same qualitative profile.
Low- and medium-degree rows occur frequently, consistent with the fact that their supports are more likely to be error-free at the channel hard-decision level.
However, the selected ensemble also contains several high-degree rows.
These rows are less likely to be error-free, but when they are reliable they influence a larger neighborhood and can rescue failure patterns not covered by smaller checks.
Thus, the gain comes from empirical rescue coverage and complementarity across boosted rows.

\section{Numerical Results}
\label{sec:results}

\begin{figure}[t]
    \centering
    \resizebox{0.98\linewidth}{!}{\input{fig/bg1_n144_k96_fer_curve.tikz}}
    \caption{FER versus \(E_\mathrm{b}/N_0\) for \(\Ccal_\mathrm{5G,BG1}(144,96)\). RBE-E32 (with $I=20$ per member) is compared against BP-20, BP-640, affine subcode, saturated-min-sum, and MBBP-LW ensembles of equal size, broken-graph AED (with $5$ ensemble members), and the ML reference.}
    \label{fig:main_fer}
\end{figure}
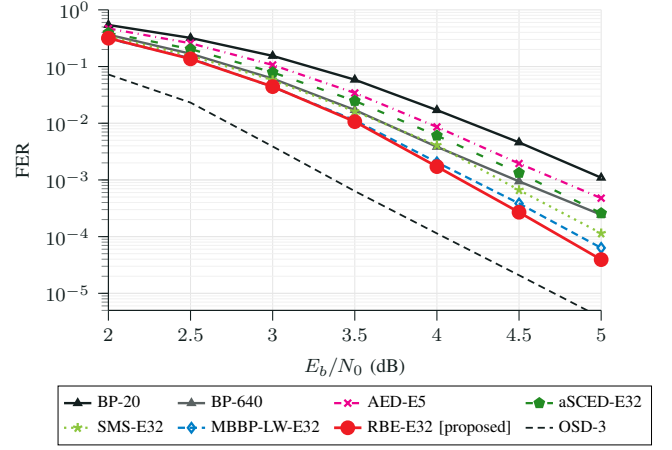

\subsection{Comparison to Other Ensemble Methods}
\label{subsec:main_benchmark}

\begin{figure*}[t]
    \centering
    \input{fig/bg1_triplet_rowboost.tikz}
    \caption{Decoding threshold versus block length for \(\Ccal_\mathrm{5G,BG1}(n,R\cdot n)\) codes at three rates. RBE uses $\beta=4$ and consistently lowers the BP threshold; the shaded region marks the gain. ML references shown where available.}
    \label{fig:bg1_triplet}
    \vspace{-1em}
\end{figure*}
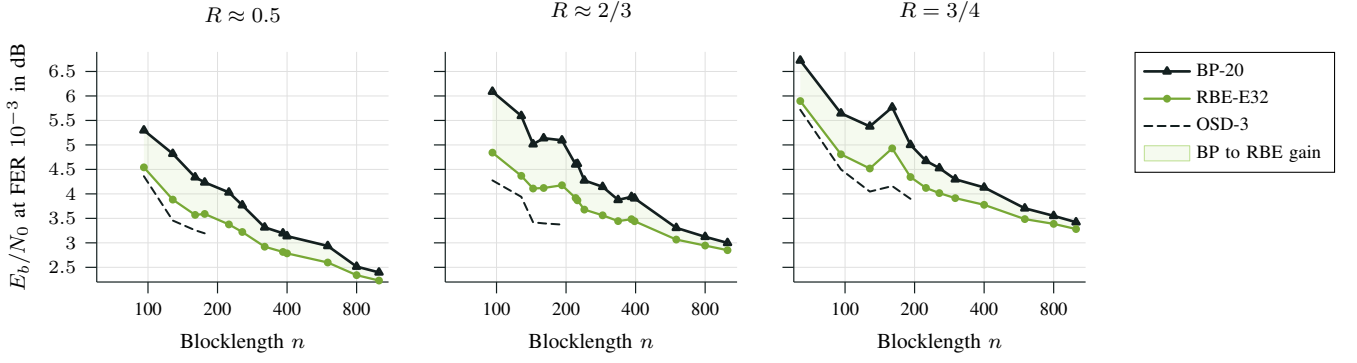

Fig.~\ref{fig:main_fer} shows the \ac{FER} on the 5G~NR BG1 \((144,96)\) code. Over the entire simulated \ac{SNR} range, RBE-E32 clearly improves on plain BP-20 by roughly an order of magnitude at moderate-to-high \ac{SNR}, and also outperforms long-iteration BP-640: additional diverse trajectories are more valuable than a longer single one. 

At the same ensemble size \(E=32\), \ac{RBE} achieves the lowest \ac{FER} among all considered ensembles, outperforming the affine subcode ensemble~\cite{mandelbaum2026affinesubcode}, the saturated-min-sum ensemble~\cite{wehn2016afterburner}, and the \ac{MBBP} ensemble~\cite{Huber,hehn2010multiple} with $5$ low weight checks appended to the original \ac{PCM}. 
Broken-graph \ac{AED} is limited to \(E=5\) members by the lifting structure of the code and, accordingly, provides only a small gain over plain BP-20. The residual gap to the \ac{ML} reference is dominated by frames whose correct codeword is absent from the candidate list; mis-selections among valid candidates, including the rare event that the gate converges to a wrong codeword, account for only a small fraction of the errors.

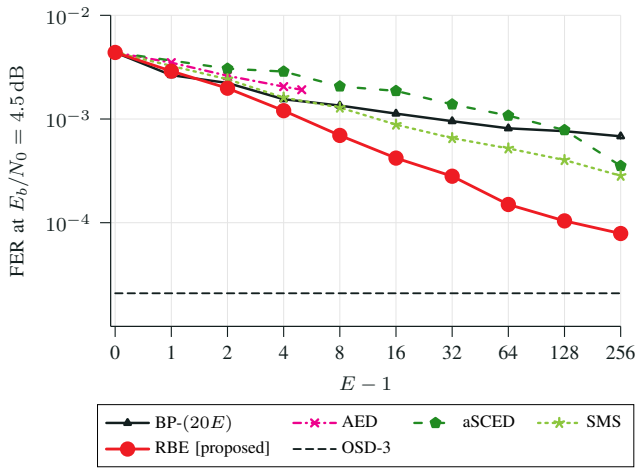
\begin{figure}[!b]
    \vspace{-1em}
    \centering
    \resizebox{0.98\linewidth}{!}{\input{fig/rbe_ensemble_size_tradeoff.tikz}}
    \caption{Scaling on \(\Ccal_\mathrm{5G,BG1}(144,96)\). FER vs.~ensemble size $E$ at \(E_\mathrm{b}/N_0=4.5\,\mathrm{dB}\): RBE vs.~ decoder (ensembles) of equal complexity.}
    \label{fig:rbe_scaling}
\end{figure}

\subsection{Ensemble-Size Scaling and Schedule Generality}
\label{subsec:ensemble_size_scaling}

Fig.~\ref{fig:rbe_scaling} compares \ac{RBE} with a single \ac{BP} decoder using the same total iteration budget of \(20E\) iterations, i.e., the same budget spent on a single trajectory. All ensembles are gated: the ensemble members are only invoked if a plain BP-20 gate does not converge to a valid codeword. The gate thereby acts as an additional member, i.e., the effective ensemble size is increased by one. \ac{RBE} consistently outperforms both the matched-budget single decoder and the affine subcode ensemble of equal size. This confirms that, for a fixed iteration budget, decoder diversity is more valuable than extending a single trajectory. Notably, \ac{RBE} continues to scale with the ensemble size: the \ac{FER} improves steadily up to \(E=256\) without showing a performance saturation.

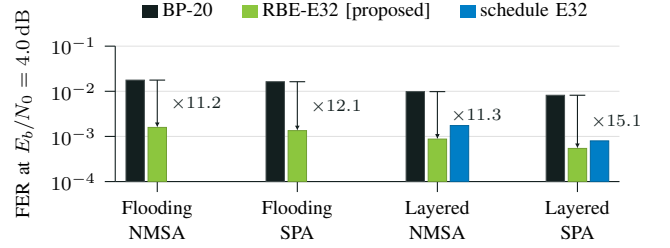
\begin{figure}[t]
    \centering
    \resizebox{0.98\linewidth}{!}{\input{fig/fig5_decoder_architecture_generality.tikz}}
    \vspace{-0.5em}
    \caption{Schedule and decoder generality on \(\Ccal_\mathrm{5G,BG1}(144,96)\). RBE-E32 versus BP-20 under flooding and layered NMSA at \(E_\mathrm{b}/N_0=4.0\,\mathrm{dB}\), with a layered schedule-diversity ensemble for reference.}
    \label{fig:decoder_architecture_generality}
    \vspace{-1.25em}
\end{figure}

Since \ac{RBE} only acts at the \ac{CN} output, it is agnostic to the underlying kernel and schedule. This is verified in Fig.~\ref{fig:decoder_architecture_generality}: under both flooding and layered \ac{NMSA}, \ac{RBE}-E32 yields nearly identical gains over the respective plain BP-20 decoder. On the layered schedule, \ac{RBE} also outperforms a schedule-diversity ensemble~\cite{krieg2025ensemble} of equal size without altering the update order. Note that, unlike this baseline, \ac{RBE} composes with any schedule, including the layered one considered here.

\subsection{Length and Rate Generality}
\label{subsec:code_generality}

Fig.~\ref{fig:bg1_triplet} reports the decoding threshold at \ac{FER} \(10^{-3}\) versus block length for three 5G~NR rate families. RBE-E32 lowers the BP-20 threshold across all considered lengths and rates. The gains are largest at short block lengths, where finite-length graph effects dominate, and gradually diminish with increasing \(n\), yet remain visible up to \(n=1000\). Note that the same row-boost construction rule (identical \(\beta\), member cardinality, and selection procedure) is applied to every code in the family; only the selected row indices adapt to each \ac{PCM}, without any per-code tuning. Larger ensemble sizes typically close the remaining gap to the \ac{ML} threshold further; for the shortest codes, RBE-E64 approaches \ac{ML} performance.

\subsection{Cross-Code Transfer}
\label{subsec:cross_code}

\begin{figure*}[t]
    \centering
    \input{fig/fig6_code_generality_plain_rbe.tikz}
    \caption{Performance gains for different channel codes evaluated at \(\gamma=E_\mathrm{b}/N_0\) using an ensemble of size $E$. \ac{RBE} (gated) is compared with plain BP-20, the matched-budget BP-($20E$), \ac{SMS} (gated) and the ML or OSD-3 reference where available. Corresponding $\mathbf{H}$ matrices are shown beneath each code.}
    \label{fig:cross_code_generality}
    \vspace{-1.5em}
\end{figure*}
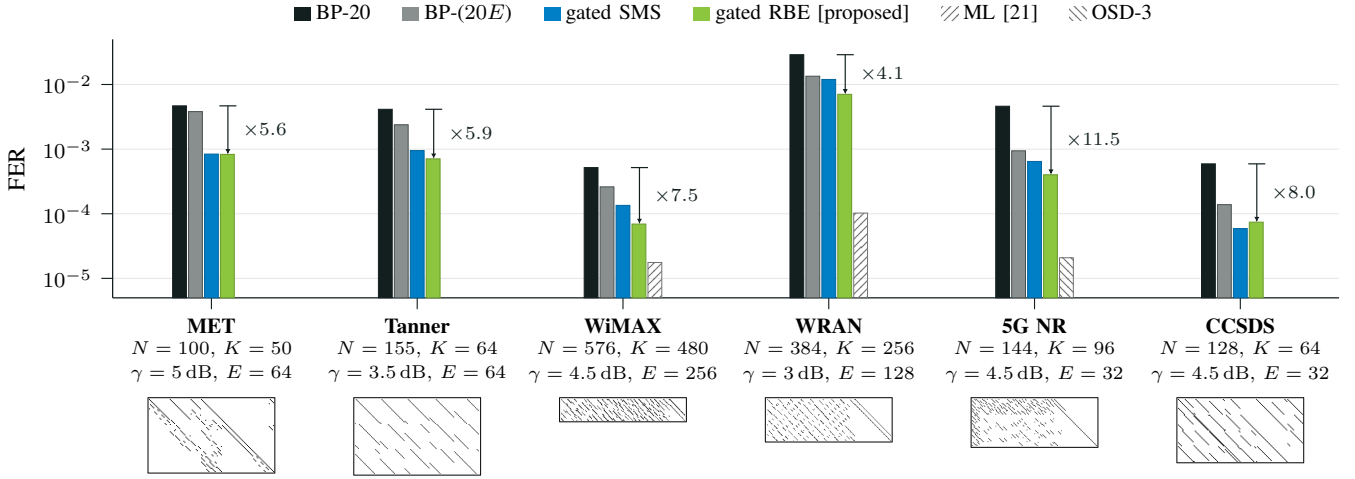

Fig.~\ref{fig:cross_code_generality} extends the comparison to five further short \ac{LDPC} codes: MET \((100,50)\)~\cite{richardson2002met}, Tanner \((155,64)\)~\cite{tanner2004circulant}, WiMAX \((576,480)\)~\cite{ieee80216e}, WRAN \((384,256)\)~\cite{ieee80222}, and CCSDS \((128,64)\)~\cite{ccsds231}. Plain gated \ac{RBE}, selected by the channel metric without any post-processing, improves over BP-20 on every code and also outperforms the matched-budget BP-(20$E$) throughout. Note that the gain is consistent across the structurally distinct parity-check matrices shown beneath the bars, none of which received per-code tuning.

\section{Discussion and Conclusion}
\label{sec:conclusion}

We introduced \ac{RBE}-\ac{BP} decoding for short \ac{LDPC} codes. \ac{RBE} preserves the parity-check matrix, the Tanner-graph storage, the decoder kernel, the syndrome validation, and the candidate metric, and creates diversity solely by boosting selected rows in different ensemble members. On the 5G~NR BG1 \((144,96)\) code, \ac{RBE} reduces the BP-20 \ac{FER} by roughly an order of magnitude and outperforms affine subcode and saturated-min-sum ensembles of equal size; larger ensembles further approach the \ac{ML} threshold for the shortest codes. The gains transfer across 5G~NR block lengths and rates as well as to other code families with the greedy row search. We found that boosted rows reliably anchor failed \ac{BP} trajectories, especially if their support is error-free. Notably, \ac{RBE} does not rely on a highly tuned boost design. Even random row-boost configurations provide measurable gains and transfer across different scenarios.

\ac{RBE} is a low-overhead mechanism for decoder-side diversity, not an \ac{ML} decoder. It helps when changing the relative strength of selected parity checks redirects a failed \ac{BP} trajectory toward the transmitted codeword; it is least effective when all boosted trajectories fail or when the received word is closer to a wrong valid codeword. The residual gap to \ac{ML} is therefore a list-completeness limit: at high \ac{SNR}, the correct codeword is predominantly absent from \(\Lcal\).
Since the selected rows transfer across the rate family without re-tuning and the useful boost factor spans a broad range, the design effort remains small. Candidate re-encoding, \ac{OSD}-0, soft-information fusion, the energy--latency optimization of the execution policy, and adaptive online row selection are deferred to future work.
\vspace{-0.5em}
\bibliographystyle{IEEEtran}
\bibliography{references}

\end{document}

%% file: corporateColours.tex
\definecolor{mittelblau}{RGB}{0, 126, 198}
\definecolor{violettblau}{cmyk}{0.9, 0.6, 0, 0}
\definecolor{rot}{RGB}{238, 28 35}
\definecolor{apfelgruen}{RGB}{140, 198, 62}
\definecolor{gelb}{RGB}{255, 229, 0}
\definecolor{orange}{RGB}{244, 111, 33}
\definecolor{pink}{RGB}{237, 0, 140}
\definecolor{lila}{RGB}{128, 10, 145}
\definecolor{hellgrau}{RGB}{224, 224, 224}
\definecolor{mittelgrau}{RGB}{128, 128, 128}
\definecolor{dunkelgrau}{RGB}{80,80,80}
\definecolor{anthrazit}{RGB}{19, 31, 31}
\definecolor{darkgreen}{RGB}{34,139,34}
\definecolor{aqua}{RGB}{0, 255, 255}

\definecolor{lightgray}{RGB}{211,211,211}

\definecolor{neuesgruen}{RGB}{61, 173, 65}
\definecolor{dunklereshellgrau}{RGB}{176, 176, 176}
\definecolor{neuesgelb}{RGB}{255,160,0}
\definecolor{neuescyan}{RGB}{69,185,224}

\definecolor{tollesgruen}{RGB}{0,217,171}
\definecolor{tollesmagenta}{RGB}{197,67,143}

\definecolor{tollesgelb}{RGB}{255,199,95}
\definecolor{tollesrot}{RGB}{255,111,145}

\colorlet{R12}{apfelgruen}
\colorlet{R23}{mittelblau}
\colorlet{R45}{pink}

%% file: fig/_rbe_dec_block.tex
\begin{tikzpicture}[
    font=\footnotesize,
    >={Latex[length=2.0mm,width=1.5mm]},
    rbblock/.style={
        draw=#1,
        line width=0.75pt,
        fill=#1!8,
        align=center,
        minimum height=8.5mm,
        minimum width=31mm,
        inner xsep=3pt,
        inner ysep=2pt
    },
    tallblock/.style={
        draw,
        align=center,
        minimum height=70mm,
        minimum width=10mm,
        inner xsep=2pt,
        inner ysep=2pt
    },
    arr/.style={-Latex, line width=0.55pt, draw=black},
    line/.style={line width=0.55pt, draw=black},
    note/.style={font=\scriptsize, align=center},
    vnode/.style={circle, draw=#1!55, line width=0.5pt, fill=#1!12, minimum size=3.6pt, inner sep=0pt},
    cnode/.style={rectangle, draw=#1!55, line width=0.5pt, fill=#1!12, minimum size=3.6pt, inner sep=0pt},
    cnodeboost/.style={rectangle, draw=#1, line width=0.9pt, fill=#1!30, minimum size=3.6pt, inner sep=0pt},
    tgedge/.style={line width=0.3pt, draw=#1!35},
    tgboost/.style={line width=1.7pt, draw=#1}
]
\def\yA{3.40}
\def\yGA{2.15}
\def\yB{0.65}
\def\yGB{-0.60}
\def\yDots{-1.55}
\def\yC{-2.75}
\def\yGC{-4.00}
\def\yMid{0.325}
\node (llr) at (0,\yMid) {\(\mathbf{L}_{\mathrm{ch}}\)};
\node[rbblock=R12] (rb1) at (2.85,\yA) {
    \(\ac{BP}\!\left(\mathbf{L}_{\mathrm{ch}};\mathbf{H},\textcolor{R12}{\mathcal{B}_1},\textcolor{R12}{\beta_1}\right)\)
};
\node[rbblock=R23] (rb2) at (2.85,\yB) {
    \(\ac{BP}\!\left(\mathbf{L}_{\mathrm{ch}};\mathbf{H},\textcolor{R23}{\mathcal{B}_2},\textcolor{R23}{\beta_2}\right)\)
};
\node[note] (dots) at (2.85,\yDots) {\(\vdots\)};
\node[rbblock=R45] (rbM) at (2.85,\yC) {
    \(\ac{BP}\!\left(\mathbf{L}_{\mathrm{ch}};\mathbf{H},\textcolor{R45}{\mathcal{B}_M},\textcolor{R45}{\beta_M}\right)\)
};
\begin{scope}[shift={(2.85,\yGA)}]
    \coordinate (g1v1) at (-1.05,0.4);
    \coordinate (g1v2) at (-0.35,0.4);
    \coordinate (g1v3) at (0.35,0.4);
    \coordinate (g1v4) at (1.05,0.4);
    \coordinate (g1c1) at (-0.7,-0.4);
    \coordinate (g1c2) at (0,-0.4);
    \coordinate (g1c3) at (0.7,-0.4);
    \draw[tgedge=R12] (g1c2) -- (g1v2);
    \draw[tgedge=R12] (g1c2) -- (g1v3);
    \draw[tgedge=R12] (g1c2) -- (g1v4);
    \draw[tgedge=R12] (g1c3) -- (g1v1);
    \draw[tgedge=R12] (g1c3) -- (g1v3);
    \draw[tgedge=R12] (g1c3) -- (g1v4);
    \draw[tgboost=R12] (g1c1) -- (g1v1);
    \draw[tgboost=R12] (g1c1) -- (g1v2);
    \draw[tgboost=R12] (g1c1) -- (g1v3);
    \node[vnode=R12] at (g1v1) {};
    \node[vnode=R12] at (g1v2) {};
    \node[vnode=R12] at (g1v3) {};
    \node[vnode=R12] at (g1v4) {};
    \node[cnode=R12] at (g1c2) {};
    \node[cnode=R12] at (g1c3) {};
    \node[cnodeboost=R12] at (g1c1) {};
\end{scope}
\begin{scope}[shift={(2.85,\yGB)}]
    \coordinate (g2v1) at (-1.05,0.4);
    \coordinate (g2v2) at (-0.35,0.4);
    \coordinate (g2v3) at (0.35,0.4);
    \coordinate (g2v4) at (1.05,0.4);
    \coordinate (g2c1) at (-0.7,-0.4);
    \coordinate (g2c2) at (0,-0.4);
    \coordinate (g2c3) at (0.7,-0.4);
    \draw[tgedge=R23] (g2c1) -- (g2v1);
    \draw[tgedge=R23] (g2c1) -- (g2v2);
    \draw[tgedge=R23] (g2c1) -- (g2v3);
    \draw[tgedge=R23] (g2c3) -- (g2v1);
    \draw[tgedge=R23] (g2c3) -- (g2v3);
    \draw[tgedge=R23] (g2c3) -- (g2v4);
    \draw[tgboost=R23] (g2c2) -- (g2v2);
    \draw[tgboost=R23] (g2c2) -- (g2v3);
    \draw[tgboost=R23] (g2c2) -- (g2v4);
    \node[vnode=R23] at (g2v1) {};
    \node[vnode=R23] at (g2v2) {};
    \node[vnode=R23] at (g2v3) {};
    \node[vnode=R23] at (g2v4) {};
    \node[cnode=R23] at (g2c1) {};
    \node[cnode=R23] at (g2c3) {};
    \node[cnodeboost=R23] at (g2c2) {};
\end{scope}
\begin{scope}[shift={(2.85,\yGC)}]
    \coordinate (g3v1) at (-1.05,0.4);
    \coordinate (g3v2) at (-0.35,0.4);
    \coordinate (g3v3) at (0.35,0.4);
    \coordinate (g3v4) at (1.05,0.4);
    \coordinate (g3c1) at (-0.7,-0.4);
    \coordinate (g3c2) at (0,-0.4);
    \coordinate (g3c3) at (0.7,-0.4);
    \draw[tgedge=R45] (g3c1) -- (g3v1);
    \draw[tgedge=R45] (g3c1) -- (g3v2);
    \draw[tgedge=R45] (g3c1) -- (g3v3);
    \draw[tgedge=R45] (g3c2) -- (g3v2);
    \draw[tgedge=R45] (g3c2) -- (g3v3);
    \draw[tgedge=R45] (g3c2) -- (g3v4);
    \draw[tgboost=R45] (g3c3) -- (g3v1);
    \draw[tgboost=R45] (g3c3) -- (g3v3);
    \draw[tgboost=R45] (g3c3) -- (g3v4);
    \node[vnode=R45] at (g3v1) {};
    \node[vnode=R45] at (g3v2) {};
    \node[vnode=R45] at (g3v3) {};
    \node[vnode=R45] at (g3v4) {};
    \node[cnode=R45] at (g3c1) {};
    \node[cnode=R45] at (g3c2) {};
    \node[cnodeboost=R45] at (g3c3) {};
\end{scope}

\node[tallblock] (cand) at (5.85,\yMid) {
    \rotatebox{90}{\begin{tabular}{c}candidate\\[-1pt]processing\end{tabular}}
};
\node[tallblock] (ml) at (7.65,\yMid) {
    \rotatebox{90}{\begin{tabular}{c}ML in\\[-1pt]the list\end{tabular}}
};
\node (out) at (9.35,\yMid) {\(\hat{\mathbf{c}}\in\mathcal{C}\)};
\coordinate (busL) at (0.85,\yMid);
\draw[line] (llr.east) -- (busL);
\draw[line] (busL |- rb1.west) -- (busL |- rbM.west);
\draw[arr] (busL |- rb1.west) -- (rb1.west);
\draw[arr] (busL |- rb2.west) -- (rb2.west);
\draw[arr] (busL |- rbM.west) -- (rbM.west);
\draw[arr] (rb1.east) -- node[above, font=\scriptsize] {\(\hat{\mathbf{x}}_1\)}
    (cand.west |- rb1.east);
\draw[arr] (rb2.east) -- node[above, font=\scriptsize] {\(\hat{\mathbf{x}}_2\)}
    (cand.west |- rb2.east);
\draw[arr] (rbM.east) -- node[below, font=\scriptsize] {\(\hat{\mathbf{x}}_M\)}
    (cand.west |- rbM.east);
\draw[arr] (cand.east) -- node[above, font=\scriptsize, midway] {\(\mathcal{L}\)}
    (ml.west);
\draw[arr] (ml.east) -- (out.west);
\end{tikzpicture}

%% file: fig/_bg1_n144_k96_rowboost_trajectory_example.tikz
\begin{tikzpicture}

\begin{axis}[
    width=\linewidth,
    height=0.45\linewidth,
    xmin=0, xmax=105,
    ymin=0, ymax=40,
    xtick={0,20,40,60,80,100},
    ytick={0,10,20,30,40},
    axis x line*=bottom,
    axis y line*=left,
    xlabel={Iteration \(i\)},
    ylabel={Syndrome weight  \(w_{\mathrm{s}}^{(i)}\)},
    label style={font=\footnotesize},
    ticklabel style={font=\scriptsize},
    grid=major,
    major grid style={draw=hellgrau, line width=0.35pt},
    axis line style={draw=anthrazit, line width=0.45pt},
    tick align=outside,
    tick style={draw=anthrazit, line width=0.45pt},
    every axis plot/.append style={
        line join=round,
        line cap=round
    },
    legend style={
        draw=black,
        fill=none,
        font=\scriptsize,
        at={(0.5,-0.45)},
        anchor=north,
        legend columns=3,
        /tikz/every even column/.append style={column sep=0.35cm}
    },
    legend cell align={left},
    clip mode=individual,
]

\addlegendimage{draw=anthrazit, very thick, mark=none}
\addlegendentry{\ac{BP}-100}

\addlegendimage{draw=R23, opacity=0.35, semithick, mark=none}
\addlegendentry{\ac{RBE}, not conv.}

\addlegendimage{
    draw=R12,
    very thick,
    mark=none,
    mark size=1.5pt,
    mark options={fill=R12, draw=white, line width=0.25pt}
}
\addlegendentry{\ac{RBE}, conv.}

\addplot[
    draw=anthrazit,
    very thick,
    mark=none
] coordinates {
    (0,14) (1,25) (2,19) (3,17) (4,11) (5,20) (6,16) (7,15)
    (8,18) (9,30) (10,17) (11,11) (12,20) (13,18) (14,21)
    (15,24) (16,23) (17,8) (18,19) (19,19) (20,21) (21,13)
    (22,12) (23,16) (24,19) (25,20) (26,25) (27,20) (28,18)
    (29,14) (30,13) (31,14) (32,16) (33,14) (34,19) (35,14)
    (36,18) (37,13) (38,14) (39,25) (40,28) (41,35) (42,17)
    (43,19) (44,15) (45,12) (46,19) (47,17) (48,14) (49,16)
    (50,12) (51,14) (52,21) (53,20) (54,25) (55,17) (56,23)
    (57,30) (58,22) (59,25) (60,38) (61,34) (62,28) (63,36)
    (64,36) (65,32) (66,33) (67,24) (68,24) (69,21) (70,24)
    (71,17) (72,14) (73,21) (74,13) (75,14) (76,10) (77,13)
    (78,16) (79,16) (80,15) (81,12) (82,16) (83,22) (84,19)
    (85,13) (86,14) (87,15) (88,14) (89,18) (90,18) (91,24)
    (92,13) (93,15) (94,12) (95,12) (96,14) (97,11) (98,19)
    (99,18) (100,28)
};

\addplot[draw=R23, opacity=0.35, semithick, mark=none, forget plot] coordinates {
    (0,14) (1,25) (2,19) (3,17) (4,26) (5,28) (6,24) (7,25)
    (8,29) (9,25) (10,23) (11,19) (12,12) (13,12) (14,12)
    (15,11) (16,13) (17,9) (18,11) (19,15) (20,21)
};

\addplot[draw=R23, opacity=0.35, semithick, mark=none, forget plot] coordinates {
    (0,14) (1,25) (2,24) (3,22) (4,12) (5,20) (6,12) (7,15)
    (8,14) (9,18) (10,17) (11,22) (12,22) (13,23) (14,27)
    (15,18) (16,16) (17,14) (18,13) (19,14) (20,26)
};

\addplot[draw=R23, opacity=0.35, semithick, mark=none, forget plot] coordinates {
    (0,14) (1,25) (2,18) (3,14) (4,17) (5,24) (6,22) (7,24)
    (8,24) (9,25) (10,31) (11,23) (12,20) (13,21) (14,26)
    (15,27) (16,37) (17,33) (18,33) (19,38) (20,31)
};

\addplot[draw=R23, opacity=0.35, semithick, mark=none, forget plot] coordinates {
    (0,14) (1,25) (2,19) (3,17) (4,17) (5,11) (6,10) (7,10)
    (8,12) (9,15) (10,16) (11,10) (12,17) (13,21) (14,21)
    (15,17) (16,12) (17,9) (18,10) (19,10) (20,19)
};

\addplot[draw=R23, opacity=0.35, semithick, mark=none, forget plot] coordinates {
    (0,14) (1,25) (2,19) (3,18) (4,14) (5,13) (6,16) (7,11)
    (8,12) (9,15) (10,18) (11,18) (12,14) (13,15) (14,10)
    (15,11) (16,11) (17,16) (18,14) (19,10) (20,10)
};

\addplot[draw=R23, opacity=0.35, semithick, mark=none, forget plot] coordinates {
    (0,14) (1,25) (2,19) (3,19) (4,38) (5,37) (6,31) (7,27)
    (8,20) (9,15) (10,12) (11,11) (12,15) (13,14) (14,12)
    (15,13) (16,14) (17,10) (18,23) (19,19) (20,14)
};

\addplot[draw=R23, opacity=0.35, semithick, mark=none, forget plot] coordinates {
    (0,14) (1,25) (2,25) (3,19) (4,26) (5,27) (6,23) (7,19)
    (8,35) (9,27) (10,24) (11,24) (12,17) (13,18) (14,11)
    (15,11) (16,6) (17,7) (18,9) (19,8) (20,8)
};

\addplot[draw=R23, opacity=0.35, semithick, mark=none, forget plot] coordinates {
    (0,14) (1,25) (2,23) (3,21) (4,11) (5,6) (6,6) (7,5)
    (8,5) (9,15) (10,22) (11,19) (12,15) (13,19) (14,14)
    (15,13) (16,10) (17,14) (18,11) (19,13) (20,19)
};

\addplot[
    draw=R12,
    very thick,
    mark=none,
    forget plot
] coordinates {
    (0,14) (1,25) (2,34) (3,34) (4,27) (5,28) (6,28) (7,24)
    (8,15) (9,9) (10,9) (11,7) (12,8) (13,8) (14,6) (15,0)
};

\addplot[
    draw=R12,
    very thick,
    mark=none,
    forget plot
] coordinates {
    (0,14) (1,25) (2,24) (3,26) (4,24) (5,28) (6,24) (7,11)
    (8,6) (9,0)
};

\draw[
    dashed,
    draw=mittelgrau,
    line width=0.65pt
] (axis cs:20,0) -- (axis cs:20,40);

\node[
    font=\scriptsize,
    text=mittelgrau!80!black,
    anchor=north west,
    fill=white,
    inner sep=1.2pt
] at (axis cs:21,39.2) {\(I_{\max}=20\)};

\end{axis}
\end{tikzpicture}

%% file: fig/fig2_tuning_selection.tikz
\begin{tikzpicture}
\begin{groupplot}[
    group style={
        group size=2 by 1,
        horizontal sep=1.05cm,
        yticklabels at=edge left
    },
    width=4.65cm,
    height=4.15cm,
    ymin=0,
    ymax=100,
    ytick={0,20,40,60,80,100},
    axis x line*=bottom,
    axis y line*=left,
    xlabel style={font=\footnotesize},
    ylabel style={font=\footnotesize},
    tick label style={font=\footnotesize},
    legend style={
        draw=black,
        fill=white,
        font=\footnotesize,
        at={(1,0.25)},
        anchor=east,
        legend columns=1,
        /tikz/every even column/.append style={column sep=0.35cm}
    },
    legend cell align={left},
    grid=major,
    major grid style={draw=hellgrau, line width=0.35pt},
    axis line style={draw=anthrazit, line width=0.45pt},
    tick align=outside,
    tick style={draw=anthrazit, line width=0.45pt},
    every axis plot/.append style={line join=round,line cap=round},
]

\nextgroupplot[
    xlabel={\(\beta\)},
    ylabel={Rescue coverage \(\rho(\mathcal{M})\) in \%},
    xmin=-0.25,
    xmax=12.6,
    xtick={0,2,4,6,8,10,12},
]
\addplot+[color=R23, mark=none, line width=0.85pt]
coordinates {(0,65.1333) (0.25,60.8667) (0.5,52.8) (0.75,40.5333) (1,0) (1.25,42.8) (1.5,57.5333) (2,72.6667) (3,85.4) (4,89.1333) (6,90.2667) (8,90.5333) (12,89.8)};
\addlegendentry{\(E=32\)}

\addplot+[color=R12, mark=none, line width=0.78pt]
coordinates {(0,49.4) (0.25,44.8) (0.5,38.1333) (0.75,28.0667) (1,0) (1.25,29.6) (1.5,41.1333) (2,54.6) (3,66.4667) (4,69.6667) (6,69.2667) (8,69.4) (12,69.4)};
\addlegendentry{\(E=8\)}

\nextgroupplot[
    xlabel={\(|\mathcal{B}_e|\)},
    xmin=-0.2,
    xmax=6.2,
    xtick={0,1,2,3,4,5,6},
]
\addplot+[color=R23, mark=none, line width=1pt]
coordinates {(0,0) (1,90.5333) (2,95.2667) (3,94) (4,89.4) (5,79.3333) (6,68.6)};

\addplot+[color=R12, mark=none, line width=0.9pt]
coordinates {(0,0) (1,69.4) (2,74.3333) (3,65.9333) (4,55.0667) (5,41.8) (6,32.4667)};

\end{groupplot}

\end{tikzpicture}
\vspace{-0.5cm}

%% file: fig/rbe_anchor_rescue_by_error_count.tikz
\begin{tikzpicture}
\begin{axis}[
    width=\linewidth,
    height=0.4\linewidth,
    ybar,
    bar width=8.5pt,
    ymin=0,
    ymax=40,
    ytick={0,10,20,30,40},
    axis x line*=bottom,
    axis y line*=left,
    ylabel style={align=center, text width=4cm},
    ylabel={Cond. rescue cov. \\ \(\tilde{\rho}_{\mathcal{M}_i}(q)\) in \%},
    xlabel={Number of channel errors in boosted check support $q_m$},
    symbolic x coords={k0,k1,k2,k3p},
    xtick=data,
    xticklabels={\(0\),\(1\),\(2\),\(\geq 3\)},
    enlarge x limits=0.25,
    ymajorgrids=true,
    major grid style={draw=hellgrau, line width=0.35pt},
    axis line style={draw=anthrazit, line width=0.45pt},
    tick align=outside,
    tick style={draw=anthrazit, line width=0.45pt},
    label style={font=\footnotesize},
    ticklabel style={font=\scriptsize},
    every axis plot/.append style={
        line join=round,
        line cap=round
    },
    legend style={
        draw=none,
        fill=none,
        font=\scriptsize,
        at={(0.5,1.15)},
        anchor=south,
        legend columns=2,
        /tikz/every even column/.append style={column sep=0.45cm}
    },
    legend cell align={left},
    legend image code/.code={
        \draw[#1, draw=none] (0cm,-0.09cm) rectangle (0.22cm,0.09cm);
    },
    nodes near coords={
    },
    point meta=y,
    every node near coord/.append style={
        font=\scriptsize,
        text=anthrazit,
        yshift=1pt
    },
]

\addplot[
    fill=R23,
    draw=R23!80!black,
    line width=0.45pt
] coordinates {
    (k0,35.8)
    (k1,4.7)
    (k2,2.1)
    (k3p,2.0)
};
\addlegendentry{$\mathcal{M}_1$: optimized rows}

\addplot[
    fill=R12 ,
    draw=R12!80!black,
    line width=0.45pt 
] coordinates {
    (k0,24.4)
    (k1,6.8)
    (k2,3.7)
    (k3p,2.0)
};
\addlegendentry{$\mathcal{M}_2$: random rows}

\end{axis}
\end{tikzpicture}

%% file: fig/bg1_n144_k96_fer_curve.tikz
\begin{tikzpicture}
\begin{axis}[
    width=\linewidth,
    height=0.68\linewidth,
    ymode=log,
    log origin=infty,
    xmin=2,
    xmax=5.00,
    ymin=5e-6,
    ymax=1,
    xtick={2,2.5,3,3.5,4,4.5,5},
    xlabel={\(E_b/N_0\) (dB)},
    ylabel={FER},
    grid=both,
    major grid style={
        draw=hellgrau!85,
        line width=0.36pt
    },
    minor grid style={
        draw=hellgrau!45,
        line width=0.22pt,
    },
    axis x line*=bottom,
    axis y line*=left,
    axis line style={draw=anthrazit, line width=0.50pt},
    tick align=outside,
    tick style={draw=anthrazit, line width=0.45pt},
    minor tick style={draw=anthrazit!55, line width=0.30pt},
    ticklabel style={font=\footnotesize, text=anthrazit},
    label style={font=\footnotesize, text=anthrazit},
    legend style={
        draw=black,
        fill=none,
        at={(0.5,-0.25)},
        anchor=north,
        legend columns=4,
        font=\scriptsize,
        inner xsep=2pt,
        /tikz/every even column/.append style={column sep=0.12cm}
    },
    legend cell align={left},
    legend image code/.code={
        \draw[mark repeat=2, mark phase=2, #1]
            plot coordinates {(0cm,0cm) (0.2cm,0cm) (0.4cm,0cm)};
    },
]

\addplot+[color=anthrazit, mark=triangle, mark size=1.5pt, solid, line width=1.0pt] coordinates {(2.00,0.54127127) (2.50,0.31990347) (3.00,0.15420484) (3.50,0.058655105) (4.00,0.017048429) (4.50,0.0045943464) (5.00,0.0010951865)};
\addlegendentry{BP-20}

\addplot+[color=anthrazit!70, mark=triangle, mark size=1.5pt, solid, line width=1.0pt] coordinates {(2.00,0.36481512) (2.50,0.16843914) (3.00,0.061272906) (3.50,0.017138416) (4.00,0.0038367147) (4.50,0.00094341116) (5.00,0.00024307034)};
\addlegendentry{BP-640}

\addplot+[color=magenta, mark=x, mark size=2pt, dash dot, line width=0.95pt, mark options={solid}] coordinates {(2.00,0.46903632) (2.50,0.25746482) (3.00,0.10665494) (3.50,0.034215478) (4.00,0.0086101113) (4.50,0.0019497164) (5.00,0.00047720332)};
\addlegendentry{AED-E5}

\addplot+[color=darkgreen, mark=pentagon*, mark size=2.0pt, loosely dashed, line width=0.95pt, mark options={solid}] coordinates {(2.00,0.39250654) (2.50,0.20330906) (3.00,0.079045321) (3.50,0.024807755) (4.00,0.0060536649) (4.50,0.001334465) (5.00,0.00025820108)};
\addlegendentry{aSCED-E32}

\addplot+[color=R12, mark=star, mark size=2.0pt, dotted, line width=0.95pt, mark options={solid}] coordinates {(2.00,0.33794175) (2.50,0.15535013) (3.00,0.057591623) (3.50,0.016156741) (4.00,0.0040903141) (4.50,0.00066388945) (5.00,0.00011441438)};
\addlegendentry{SMS-E32}

\addplot+[color=R23, mark=diamond, mark size=2pt, densely dashed, line width=0.95pt, mark options={solid}] coordinates {(2.00,0.30916639) (2.50,0.14023642) (3.00,0.043684555) (3.50,0.011268815) (4.00,0.0020451571) (4.50,0.00038486138) (5.00,6.3242549e-05)};
\addlegendentry{MBBP-LW-E32}

\addplot+[color=rot, mark=*, mark size=2.5pt, solid, line width=1.15pt, mark options={solid}] coordinates {(2.00,0.31630399) (2.50,0.13657559) (3.00,0.044257199) (3.50,0.01067572) (4.00,0.0017179319) (4.50,0.00026916906) (5.00,3.9337471e-05)};
\addlegendentry{RBE-E32 [proposed]}

\addplot+[color=anthrazit, mark=none, line width=0.75pt, densely dashed]
  coordinates {
      (2.00,7.2395833e-02)
      (2.50,2.3125000e-02)
      (3.00,3.8966049e-03)
      (3.50,6.3516260e-04)
      (4.00,1.1396791e-04)
      (4.50,2.0858816e-05)
      (5.00,3.7311637e-06)
  };
  \addlegendentry{OSD-3}
  
\end{axis}
\end{tikzpicture}

%% file: fig/bg1_triplet_rowboost.tikz
\begin{tikzpicture}
\begin{groupplot}[
    group style={
        group size=3 by 1,
        horizontal sep=0.75cm,
        yticklabels at=edge left
    },
    width=0.30\linewidth,
    height=0.255\linewidth,
    xmode=log,
    log basis x=2,
    xmin=60,
    xmax=1100,
    xtick={100,200,400,800},
    xticklabels={100,200,400,800},
    ymin=2.2,
    ymax=6.9,
    axis x line*=bottom,
    axis y line*=left,
    ytick={2.5,3.0,3.5,4.0,4.5,5.0,5.5,6.0,6.5},
    xlabel={Blocklength \(n\)},
    grid=major,
    major grid style={draw=hellgrau, line width=0.35pt},
    axis line style={draw=anthrazit, line width=0.45pt},
    tick align=outside,
    tick style={draw=anthrazit, line width=0.45pt},
    label style={font=\footnotesize},
    ticklabel style={font=\scriptsize},
    title style={font=\footnotesize, at={(0.5,1.02)}, anchor=south},
    every axis plot/.append style={
        line join=round,
        line cap=round
    },
    legend style={
        draw=black,
        fill=none,
        at={(3.9,1)},
        anchor=north,
        legend columns=1,
        font=\scriptsize,
        /tikz/every even column/.append style={column sep=0.38cm}
    },
    legend cell align={left},
]

\nextgroupplot[
    title={\(R\approx 0.5\)},
    ylabel={\(E_b/N_0\) at FER \(10^{-3}\) in dB},
]
\addplot[name path=bpA, color=anthrazit, mark=triangle, mark size=1.5pt, solid, line width=1.0pt]
coordinates {
    (96,5.298413) (128,4.819729) (160,4.339991) (176,4.234810)
    (224,4.030663) (256,3.769116) (320,3.316365) (384,3.195768)
    (400,3.137013) (600,2.936628) (800,2.514890) (1000,2.396452)
};
\addlegendentry{BP-20}
\addplot[name path=rbA, R12!85!black, mark=*, mark size=1.05pt, line width=0.85pt]
  coordinates {
      (96,4.541771) (128,3.883333) (160,3.570499) (176,3.590439)
      (224,3.376257) (256,3.221660) (320,2.921341) (384,2.814147)
      (400,2.784904) (600,2.598492) (800,2.340206) (1000,2.230014)
  };
  \addlegendentry{RBE-E32}
\addplot[color=anthrazit, mark=none, line width=0.75pt, densely dashed]
  coordinates {
      (96,4.360993) (128,3.455962) (160,3.257129) (176,3.194389)
  };
\addlegendentry{OSD-3}
\addplot[R12!55, fill=R12!50, fill opacity=0.16, draw=none]
fill between[of=bpA and rbA];
\addlegendimage{area legend, fill=R12!30, draw=none, fill opacity=0.16}
\addlegendentry{BP to RBE gain}

\nextgroupplot[
    title={\(R\approx 2/3\)},
]
\addplot[name path=bpB, color=anthrazit, mark=triangle, mark size=1.5pt, solid, line width=1.0pt]
coordinates {
    (96,6.088771) (128,5.593271) (144,5.014300) (160,5.137591)
    (192,5.094618) (220,4.601554) (224,4.614382) (240,4.274776)
    (288,4.144987) (336,3.877453) (384,3.939934) (396,3.909396)
    (600,3.306270) (800,3.123545) (1000,2.998862)
};
\addplot[name path=rbB, R12!85!black, mark=*, mark size=1.05pt, line width=0.85pt]
  coordinates {
      (96,4.842632) (128,4.368590) (144,4.109313) (160,4.120190)
      (192,4.176128) (220,3.916259) (224,3.868634) (240,3.679036)
      (288,3.562348) (336,3.444031) (384,3.483028) (396,3.441684)
      (600,3.066933) (800,2.944700) (1000,2.850166)
  };
\addplot[color=anthrazit, mark=none, line width=0.75pt, densely dashed]
  coordinates {
      (96,4.275896) (128,3.940183) (144,3.419066)
      (160,3.397764) (192,3.373009)
  };
\addplot[R12!55, fill=R12!50, fill opacity=0.16, draw=none]
fill between[of=bpB and rbB];

\nextgroupplot[
    title={\(R=3/4\)},
]
\addplot[name path=bpC, color=anthrazit, mark=triangle, mark size=1.5pt, solid, line width=1.0pt]
coordinates {
    (64,6.724789) (96,5.645625) (128,5.378016) (160,5.767886)
    (192,5.001612) (224,4.674473) (256,4.526773) (300,4.297886)
    (400,4.131347) (600,3.705373) (800,3.551331) (1000,3.421474)
};
\addplot[name path=rbC, R12!85!black, mark=*, mark size=1.05pt, line width=0.85pt]
  coordinates {
      (64,5.896192) (96,4.808463) (128,4.518259) (160,4.928915)
      (192,4.344203) (224,4.121501) (256,4.015889) (300,3.913151)
      (400,3.776829) (600,3.484929) (800,3.387516) (1000,3.281842)
  };
\addplot[color=anthrazit, mark=none, line width=0.75pt, densely dashed]
  coordinates {
      (64,5.717839) (96,4.499149) (128,4.046976)
      (160,4.159695) (192,3.898871)
  };
\addplot[R12!55, fill=R12!50, fill opacity=0.16, draw=none]
fill between[of=bpC and rbC];

\end{groupplot}
\end{tikzpicture}

%% file: fig/rbe_ensemble_size_tradeoff.tikz
\begin{tikzpicture}
\begin{axis}[
    width=\linewidth,
    height=0.68\linewidth,
    ymode=log,
    xmin=-0.08,
    xmax=9,
    ymin=1e-5,
    ymax=1e-2,
    xtick={0,1,2,3,4,5,6,7,8,9},
    xticklabels={0,1,2,4,8,16,32,64,128,256,512,1024},
    ytick={1e-2,1e-3,1e-4},
    xlabel={\(E-1\)},
    ylabel={FER at \(E_b/N_0=4.5\,\mathrm{dB}\)},
    grid=both,
    major grid style={
        draw=hellgrau!85,
        line width=0.36pt
    },
    minor grid style={
        draw=hellgrau!45,
        line width=0.22pt,
        densely dotted
    },
    axis x line*=bottom,
    axis y line*=left,
    axis line style={draw=anthrazit, line width=0.50pt},
    tick align=outside,
    tick style={draw=anthrazit, line width=0.45pt},
    minor tick style={draw=anthrazit!55, line width=0.30pt},
    ticklabel style={font=\footnotesize, text=anthrazit},
    label style={font=\footnotesize, text=anthrazit},
    legend style={
        draw=black,
        fill=none,
        at={(0.5,-0.25)},
        anchor=north,
        legend columns=4,
        font=\scriptsize,
        /tikz/every even column/.append style={column sep=0.24cm}
    },
    legend cell align={left},
    every axis plot/.append style={line join=round,line cap=round},
]
\addplot+[color=anthrazit, mark=triangle, mark size=1.5pt, solid, line width=1.0pt] coordinates {(0,0.0043848168) (1,0.0026484784) (2,0.002225992) (3,0.0015512895) (4,0.0013484842) (5,0.0011276001) (6,0.00095313832) (7,0.00081325069) (8,0.00076352531) (9,0.00068127199) (9.002815,0.00065189382) (10.001408,0.00054250482)};
\addlegendentry{BP-\((20E)\)}

\addplot+[color=magenta, mark=x, mark size=2pt, dash dot, line width=0.95pt, mark options={solid}] coordinates {(0,0.0043848168) (1,0.0034972186) (2,0.0026126882) (3,0.0020533377) (3.321928,0.0019150107)};
\addlegendentry{AED}

\addplot+[color=darkgreen, mark=pentagon*, mark size=2.0pt, loosely dashed, line width=0.95pt, mark options={solid}] coordinates {(0,0.0043848168) (2,0.0030618923) (3,0.0028686736) (4,0.0020656086) (5,0.0018666706) (6,0.0013852531) (7,0.0010785513) (8,0.00078256199) (9,0.00035309549)};
\addlegendentry{aSCED}

\addplot+[color=R12, mark=star, mark size=2.0pt, dotted, line width=0.95pt, mark options={solid}] coordinates {(0,0.0043848168) (2,0.0024173757) (3,0.0016098308) (4,0.001288449) (5,0.00088282613) (6,0.00065445026) (7,0.00052049247) (8,0.00040195202) (9,0.00028355826)};
\addlegendentry{SMS}

\addplot+[color=rot, mark=*, mark size=2.5pt, solid, line width=1.15pt, mark options={solid}] coordinates {(0,0.0043848168) (1,0.0028959424) (2,0.0019828285) (3,0.0012030336) (4,0.00069466013) (5,0.00041988327) (6,0.00028071882) (7,0.00014982836) (8,0.00010407924) (9,7.8659887e-05)};
\addlegendentry{RBE [proposed]}

\addplot+[color=anthrazit, mark=none, line width=0.75pt, densely dashed] coordinates {(0,2.0858816e-05) (10.101408,2.0858816e-05)};
\addlegendentry{OSD-3}

\end{axis}
\end{tikzpicture}

%% file: fig/fig5_decoder_architecture_generality.tikz
\begin{tikzpicture}
\begin{axis}[
    width=\linewidth,
    height=0.4\linewidth,
    ymode=log,
    log origin=infty,
    ymin=0.0001,
    ymax=0.1,
    ybar,
    bar width=7.5pt,
    symbolic x coords={flood_nmsa,flood_spa,layered_nmsa,layered_spa},
    xtick=data,
    xticklabels={
        {\shortstack{Flooding\\NMSA}},
        {\shortstack{Flooding\\SPA}},
        {\shortstack{Layered\\NMSA}},
        {\shortstack{Layered\\SPA}}
    },
    ylabel={FER at \(E_b/N_0=4.0\,\mathrm{dB}\)},
    ytick={1e-1,1e-2,1e-3,1e-4},
    ymajorgrids=true,
    major grid style={draw=hellgrau, line width=0.35pt},
    axis x line*=bottom,
    axis y line*=left,
    axis line style={draw=anthrazit, line width=0.45pt},
    tick align=outside,
    tick style={draw=anthrazit, line width=0.45pt},
    ticklabel style={font=\footnotesize},
    label style={font=\footnotesize},
    legend style={
        draw=none,
        fill=none,
        at={(0.5,1.12)},
        anchor=south,
        legend columns=3,
        font=\footnotesize,
        /tikz/every even column/.append style={column sep=0.35cm}
    },
    legend cell align={left},
    legend image code/.code={
        \draw[#1, draw=none] (0cm,-0.09cm) rectangle (0.22cm,0.09cm);
    },
    clip=false,
]
\addplot+[ybar, bar shift=-9pt, fill=anthrazit, draw=anthrazit!80, line width=0.35pt]
coordinates {
    (flood_nmsa,0.017702908)
    (flood_spa,0.016308594)
    (layered_nmsa,0.0098361545)
    (layered_spa,0.0081759983)
};
\addlegendentry{BP-20}
\addplot+[ybar, bar shift=0pt, fill=R12, draw=R12!80!black, line width=0.35pt]
coordinates {
    (flood_nmsa,0.0015842014)
    (flood_spa,0.0013454861)
    (layered_nmsa,0.0008734809)
    (layered_spa,0.00054253472)
};
\addlegendentry{RBE-E32 [proposed]}
\addplot+[ybar, bar shift=9pt, fill=R23, draw=R23!80!black, line width=0.35pt]
coordinates {
    (layered_nmsa,0.0017415365)
    (layered_spa,0.00079752604)
};
\addlegendentry{schedule E32}

\draw[anthrazit, line width=0.45pt]
    ([xshift=-3.2pt]axis cs:flood_nmsa,0.017702908)
    -- ([xshift=3.2pt]axis cs:flood_nmsa,0.017702908);
\draw[anthrazit, -{Latex[length=2.0pt,width=2.0pt]}, line width=0.45pt]
    (axis cs:flood_nmsa,0.017702908)
    -- node[midway, right, xshift=2pt, font=\scriptsize, text=anthrazit]
        {\(\times 11.2\)}
    (axis cs:flood_nmsa,0.0015842014);

\draw[anthrazit, line width=0.45pt]
    ([xshift=-3.2pt]axis cs:flood_spa,0.016308594)
    -- ([xshift=3.2pt]axis cs:flood_spa,0.016308594);
\draw[anthrazit, -{Latex[length=2.0pt,width=2.0pt]}, line width=0.45pt]
    (axis cs:flood_spa,0.016308594)
    -- node[midway, right, xshift=2pt, font=\scriptsize, text=anthrazit]
        {\(\times 12.1\)}
    (axis cs:flood_spa,0.0013454861);

\draw[anthrazit, line width=0.45pt]
    ([xshift=-3.2pt]axis cs:layered_nmsa,0.0098361545)
    -- ([xshift=3.2pt]axis cs:layered_nmsa,0.0098361545);
\draw[anthrazit, -{Latex[length=2.0pt,width=2.0pt]}, line width=0.45pt]
    (axis cs:layered_nmsa,0.0098361545)
    -- node[midway, right, xshift=2pt, font=\scriptsize, text=anthrazit]
        {\(\times 11.3\)}
    (axis cs:layered_nmsa,0.0008734809);

\draw[anthrazit, line width=0.45pt]
    ([xshift=-3.2pt]axis cs:layered_spa,0.0081759983)
    -- ([xshift=3.2pt]axis cs:layered_spa,0.0081759983);
\draw[anthrazit, -{Latex[length=2.0pt,width=2.0pt]}, line width=0.45pt]
    (axis cs:layered_spa,0.0081759983)
    -- node[midway, right, xshift=2pt, font=\scriptsize, text=anthrazit]
        {\(\times 15.1\)}
    (axis cs:layered_spa,0.00054253472);
\end{axis}
\end{tikzpicture}

%% file: fig/fig6_code_generality_plain_rbe.tikz
\newcommand{\pcmthumb}[1]{%
  {\setlength{\fboxsep}{0pt}\setlength{\fboxrule}{0.25pt}%
  \fbox{\includegraphics[width=0.092\linewidth]{#1}}}%
}
\begin{tikzpicture}[font=\small]
\begin{axis}[
    width=17.8cm,
    height=5cm,
    ymode=log,
    log origin=infty,
    ymin=5e-6,
    ymax=5e-2,
    xmin=-0.48,
    xmax=5.48,
    ylabel={FER},
    ybar,
    bar width=5.2pt,
    xtick={0,1,2,3,4,5},
    xticklabels={ {\shortstack{\textbf{MET}\\\scriptsize \(N=100,\,K=50\)\\\scriptsize \(\gamma=5\,\mathrm{dB},\,E=64\)}},
      {\shortstack{\textbf{Tanner}\\\scriptsize \(N=155,\,K=64\)\\\scriptsize \(\gamma=3.5\,\mathrm{dB},\,E=64\)}},
      {\shortstack{\textbf{WiMAX}\\\scriptsize \(N=576,\,K=480\)\\\scriptsize \(\gamma=4.5\,\mathrm{dB},\,E=256\)}},
      {\shortstack{\textbf{WRAN}\\\scriptsize \(N=384,\,K=256\)\\\scriptsize \(\gamma=3\,\mathrm{dB},\,E=128\)}},
      {\shortstack{\textbf{5G NR}\\\scriptsize \(N=144,\,K=96\)\\\scriptsize \(\gamma=4.5\,\mathrm{dB},\,E=32\)}},
      {\shortstack{\textbf{CCSDS}\\\scriptsize \(N=128,\,K=64\)\\\scriptsize \(\gamma=4.5\,\mathrm{dB},\,E=32\)}} },
    xticklabel style={align=center, font=\footnotesize, text width=2.55cm},
    ticklabel style={font=\footnotesize},
    ytick={1e-1,1e-2,1e-3,1e-4,1e-5,1e-6,1e-7},
    axis x line*=bottom,
    axis y line*=left,
    axis line style={draw=anthrazit, line width=0.48pt},
    tick align=outside,
    tick style={draw=anthrazit, line width=0.42pt},
    ymajorgrids=true,
    major grid style={draw=hellgrau!80, line width=0.30pt},
    minor y tick num=9,
    legend style={
        draw=none,
        fill=none,
        at={(0.5,1.02)},
        anchor=south,
        legend columns=6,
        font=\footnotesize,
        /tikz/every even column/.append style={column sep=0.35cm}
    },
    legend cell align={left},
    legend image code/.code={
        \draw[#1, draw=none] (0cm,-0.09cm) rectangle (0.22cm,0.09cm);
    },
    clip=false,
]

\addplot+[ybar, bar shift=-12pt, fill=anthrazit, draw=anthrazit, line width=0.35pt]
coordinates { (0,0.00467289719626) (1,0.00413380603278) (2,0.000518120659722) (3,0.0289103190104) (4,0.0046111462163) (5,0.000591671408296) };
\addlegendentry{BP-20}
\addplot+[ybar, bar shift=-6pt, fill=anthrazit!50, draw=anthrazit!78, line width=0.28pt]
coordinates { (0,0.0038058557243) (1,0.00238355302133) (2,0.000260755750868) (3,0.013437906901) (4,0.000941856234681) (5,0.000138353013758) };
\addlegendentry{BP-($20E$)}
\addplot+[ybar, bar shift=0pt, fill=R23, draw=R23!80!black, line width=0.28pt]
coordinates {(0,0.0008382161458) (1,0.0009494357639) (2,0.0001347442794) (3,0.01196289062) (4,0.0006447816506) (5,5.897116546e-05)};
\addlegendentry{gated SMS}
\addplot+[ybar, bar shift=6pt, fill=R12, draw=R12!80!black, line width=0.42pt]
coordinates { (0,0.000827492211838) (1,0.000706195078002) (2,6.91731770833e-05) (3,0.00702921549479) (4,0.000400917202819) (5,7.40213597074e-05) };
\addlegendentry{gated RBE [proposed]}
\addplot+[
    ybar,
    bar shift=12pt,
    fill=black!12,
    draw=black!55,
    pattern=north east lines,
    pattern color=black!55,
    line width=0.30pt
]
coordinates { (2,1.761e-05) (3,0.0001025) };
\addlegendentry{ML \cite{channelcodes}}

\addplot+[
    ybar,
    bar shift=12pt,
    fill=black!12,
    draw=black!55,
    pattern=north west lines,
    pattern color=black!55,
    line width=0.30pt
]
coordinates { (4,2.0858816e-05) };
\addlegendentry{OSD-3}

\draw[anthrazit, line width=0.45pt]
    ([xshift=6.2pt-3.2pt]axis cs:0,0.00467289719626)
    -- ([xshift=6.2pt+3.2pt]axis cs:0,0.00467289719626);

\draw[anthrazit, -{Latex[length=2.0pt,width=2.0pt]}, line width=0.45pt]
    ([xshift=6.2pt]axis cs:0,0.00467289719626)
    -- node[midway, right, xshift=2pt, font=\scriptsize, text=anthrazit]
        {\(\times 5.6\)}
    ([xshift=6.2pt]axis cs:0,0.000827492211838);

\draw[anthrazit, line width=0.45pt]
    ([xshift=6.2pt-3.2pt]axis cs:1,0.00413380603278)
    -- ([xshift=6.2pt+3.2pt]axis cs:1,0.00413380603278);

\draw[anthrazit, -{Latex[length=2.0pt,width=2.0pt]}, line width=0.45pt]
    ([xshift=6.2pt]axis cs:1,0.00413380603278)
    -- node[midway, right, xshift=2pt, font=\scriptsize, text=anthrazit]
        {\(\times 5.9\)}
    ([xshift=6.2pt]axis cs:1,0.000706195078002);

\draw[anthrazit, line width=0.45pt]
    ([xshift=6.2pt-3.2pt]axis cs:2,0.000518120659722)
    -- ([xshift=6.2pt+3.2pt]axis cs:2,0.000518120659722);

\draw[anthrazit, -{Latex[length=2.0pt,width=2.0pt]}, line width=0.45pt]
    ([xshift=6.2pt]axis cs:2,0.000518120659722)
    -- node[midway, right, xshift=2pt, font=\scriptsize, text=anthrazit]
        {\(\times 7.5\)}
    ([xshift=6.2pt]axis cs:2,6.91731770833e-05);

\draw[anthrazit, line width=0.45pt]
    ([xshift=6.2pt-3.2pt]axis cs:3,0.0289103190104)
    -- ([xshift=6.2pt+3.2pt]axis cs:3,0.0289103190104);

\draw[anthrazit, -{Latex[length=2.0pt,width=2.0pt]}, line width=0.45pt]
    ([xshift=6.2pt]axis cs:3,0.0289103190104)
    -- node[midway, right, xshift=2pt, font=\scriptsize, text=anthrazit]
        {\(\times 4.1\)}
    ([xshift=6.2pt]axis cs:3,0.00702921549479);

\draw[anthrazit, line width=0.45pt]
    ([xshift=6.2pt-3.2pt]axis cs:4,0.0046111462163)
    -- ([xshift=6.2pt+3.2pt]axis cs:4,0.0046111462163);

\draw[anthrazit, -{Latex[length=2.0pt,width=2.0pt]}, line width=0.45pt]
    ([xshift=6.2pt]axis cs:4,0.0046111462163)
    -- node[midway, right, xshift=2pt, font=\scriptsize, text=anthrazit]
        {\(\times 11.5\)}
    ([xshift=6.2pt]axis cs:4,0.000400917202819);

\draw[anthrazit, line width=0.45pt]
    ([xshift=6.2pt-3.2pt]axis cs:5,0.000591671408296)
    -- ([xshift=6.2pt+3.2pt]axis cs:5,0.000591671408296);

\draw[anthrazit, -{Latex[length=2.0pt,width=2.0pt]}, line width=0.45pt]
    ([xshift=6.2pt]axis cs:5,0.000591671408296)
    -- node[midway, right, xshift=2pt, font=\scriptsize, text=anthrazit]
        {\(\times 8.0\)}
    ([xshift=6.2pt]axis cs:5,7.40213597074e-05);

\node[anchor=north, inner sep=0pt] at ([yshift=+0.40cm]axis cs:0,5e-8) {\pcmthumb{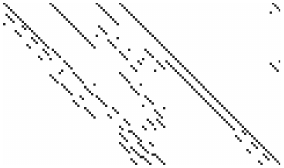}};
\node[anchor=north, inner sep=0pt] at ([yshift=+0.40cm]axis cs:1,5e-8) {\pcmthumb{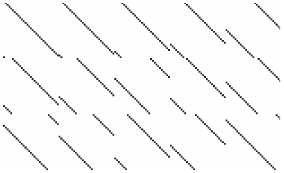}};
\node[anchor=north, inner sep=0pt] at ([yshift=+0.40cm]axis cs:2,5e-8) {\pcmthumb{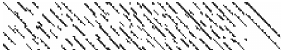}};
\node[anchor=north, inner sep=0pt] at ([yshift=+0.40cm]axis cs:3,5e-8) {\pcmthumb{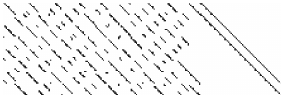}};
\node[anchor=north, inner sep=0pt] at ([yshift=+0.40cm]axis cs:4,5e-8) {\pcmthumb{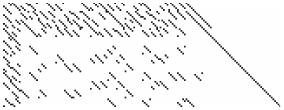}};
\node[anchor=north, inner sep=0pt] at ([yshift=+0.40cm]axis cs:5,5e-8) {\pcmthumb{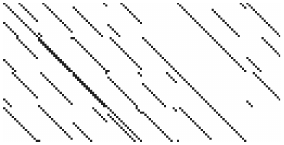}};

\end{axis}
\end{tikzpicture}